\documentclass{aa}  
\usepackage{graphicx,url}
\usepackage[normalem]{ulem}
\usepackage{mathrsfs,amssymb,amsmath}
\usepackage[usenames]{color}
\usepackage{subfigure}
\usepackage[varg]{txfonts}
\usepackage{booktabs}

\usepackage{bm}

\newcommand{\be}{\begin{equation}}
\newcommand{\ee}{\end{equation}}
\newcommand{\tP}{\widetilde P}
\newcommand{\ud}{{\rm d}}
\newcommand{\nh}{$n$}%{$n_{\rm H2}$}
\newcommand{\ellperp}{\ell_{\perp}}
\newcommand{\kms}{km~s$^{-1}$}

\begin{document}

   \title{The double signature of local cosmic-ray acceleration in star-forming regions}

%   \subtitle{}

   \author{Marco Padovani
          \inst{1}
          \and
          Alexandre Marcowith\inst{2}
          \and
          Daniele Galli\inst{1}
          \and
          Leslie K. Hunt\inst{1}
          \and
          Francesco Fontani\inst{1}
          }

   \institute{INAF--Osservatorio Astrofisico di Arcetri, 
              Largo E. Fermi 5, 50125 Firenze, Italy\\
              \email{marco.padovani@inaf.it}
         \and
             Laboratoire Univers et Particules de Montpellier, UMR 5299 du CNRS, Universit\'e de Montpellier, place E. Bataillon, cc072, 34095 Montpellier, France
             }

%   \date{}

% \abstract{}{}{}{}{} 
% 5 {} token are mandatory
 
  \abstract
   {Recently, there has been an increased interest in the study of the generation of low-energy cosmic rays ($< 1$~TeV) in shocks situated on the surface of a protostar or along protostellar jets. These locally accelerated cosmic rays offer an attractive explanation for the high levels of non-thermal emission and ionisation rates observed close to these sources.}
   {The high ionisation rate observed in some protostellar sources is generally attributed to 
   shock-generated UV photons. The aim of this article is to show that when synchrotron emission and a high ionisation rate
   are measured in the same spatial region, a locally shock-accelerated cosmic-ray flux 
   is sufficient to explain both phenomena.}
   {We assume that relativistic protons and electrons are accelerated according to the first-order Fermi acceleration mechanism, and we calculate their emerging fluxes at the shock surface. These fluxes 
     are used to compute the ionisation rate and
   the non-thermal emission at centimetre wavelengths. We then apply our model to the star-forming region 
   OMC-2 FIR 3/FIR 4. 
   Using a Bayesian analysis, we constrain the parameters of the model and estimate
   the spectral indices of the non-thermal radio emission, 
   the intensity of the magnetic field, and its degree of turbulence.}
   {We demonstrate that the local cosmic-ray acceleration model makes it possible to simultaneously explain the synchrotron emission
   along the HOPS 370 jet within the FIR 3 region
    and the ionisation rate observed near the FIR 4 protocluster.
    In particular, our model constrains the magnetic field strength ($\sim$250$-$450$~\mu$G), 
    its turbulent component ($\sim$20$-$40$~\mu$G), 
    and
    the jet velocity in the shock reference frame for the three 
    non-thermal sources of the HOPS 370 jet
    (between 350~km~s$^{-1}$ and 1000~km~s$^{-1}$). 
    }
   {Beyond the modelling of the OMC-2 FIR 3/FIR 4 system, we show how the combination of continuum
   observations at centimetre 
   wavelengths and molecular transitions is a powerful new tool for the analysis of star-forming regions: 
   These two types of observations can be simultaneously interpreted by invoking only the presence of locally accelerated cosmic 
   rays, without having to resort to shock-generated UV photons.}


   \maketitle
%
%-------------------------------------------------------------------

\section{Introduction}

In recent years there has been a renewed interest in the role of cosmic rays in the various processes of star formation (see~\citealt{Padovani+2020} for a comprehensive review).
Observational campaigns from radio to infrared frequencies and detailed theoretical models have shown how low-energy cosmic rays ($< 1$~TeV) play a key role in determining the chemical composition of the interstellar medium and the dynamic evolution of molecular clouds, from diffuse cloud scales to protostellar discs and jets.

Thanks to more and more powerful observing facilities -- such as 
the Near InfraRed Spectrograph (NIRSpec) at Keck Observatory, 
the InfraRed Camera and Spectrograph (IRCS) at the Subaru Telescope, 
the Cryogenic high-resolution InfraRed Echelle Spectrograph (CRIRES) at the Very Large Telescope, 
the United Kingdom InfraRed Telescope (UKIRT), and 
the Heterodyne Instrument for the Far Infrared (HIFI) at Herschel -- new methods have been developed to determine the influence of cosmic rays in diffuse regions of molecular clouds. It has been shown that the abundance of molecular ions such as H$_3^+$, OH$^+$, and H$_2$O$^+$ is determined by the flux of Galactic cosmic rays. It has been possible to constrain the ionisation rate of cosmic rays, $\zeta$, which reaches values of up to $\sim10^{-15}$~s$^{-1}$ \citep{IndrioloMcCall2012,Porras+2014,Indriolo+2015,Zhao+2015,Bacalla+2019}.
Observations of HCO$^+$, DCO$^+$, and CO with the 
Institute for Radio Astronomy in the Millimetre range (IRAM-30m)
telescope showed that the ionisation rate decreases at higher densities typical of starless cores ($n>10^4$~cm$^{-3}$) despite a two orders of magnitude spread
($5\times10^{-18}\lesssim\zeta/$s$^{-1}\lesssim4\times10^{-16}$; \citealt{Caselli+1998,MaretBergin2007,Fuente+2016}).
Such a spread 
could be due both to uncertainties in the chemical models and the configuration of the magnetic field lines \citep{PadovaniGalli2011,Padovani+2013,Silsbee+2018}.
Recently, \citet{Bovino+2020} developed a new method for the calculation of $\zeta$ based on H$_2$D$^+$ and other isotopologues of H$_3^+$. This method has been applied to observations carried out with the
Atacama Pathfinder EXperiment (APEX) and IRAM-30m telescopes for a large sample of massive star-forming regions, obtaining $7\times10^{-18}\lesssim\zeta/$s$^{-1}\lesssim6\times10^{-17}$ \citep{Sabatini+2020}.

The clear decrease in $\zeta$ with increasing density has been explained in a quantitative way by theoretical models: While propagating in a molecular cloud, cosmic rays lose energy due to collisions with molecular hydrogen and heavier species \citep{Padovani+2009,Padovani+2018a}. However, recent observations have estimated an unexpectedly very high ionisation rate in protostellar environments, where the Galactic cosmic-ray flux is strongly attenuated due to the high densities. For example,
in the knot B1 along the protostellar jet of the Class 0 source L1157, \citet{Podio+2014} estimated $\zeta=3\times10^{-16}$~s$^{-1}$, and in the intermediate mass star-forming region OMC-2 FIR 4, three different studies using different molecular tracers (N$_2$H$^+$, HCO$^+$, HC$_3$N, HC$_5$N, and c-C$_3$H$_2$) and instruments (Herschel and 
the NOrthern Extended Millimetre Array, NOEMA) estimated the same value of $\zeta=4\times10^{-14}$~s$^{-1}$ \citep{Ceccarelli+2014,Fontani+2017,Favre+2018}.

\citet{Padovani+2015,Padovani+2016} have shown that it is possible that a flux of cosmic rays sufficient to explain the observational estimates of enhanced ionisation rates can be generated locally, in shocks along the protostellar jet or on the surface of the protostar itself. However, in some cases, shock-generated UV photons could explain the overabundance of some molecular species \citep{Karska+2018}. An additional method is required to discriminate between these two mechanisms.

A peculiar feature of cosmic rays, in particular of the electronic component, is the synchrotron emission generated during the spiral motion of electrons around the magnetic field lines. To date, there have been many non-thermal emission detections in protostellar jets, both low-mass \citep{Ainsworth+2014} and high-mass \citep[e.g.][]{Beltran+2016,Rodriguez-Kamenetzky+2017,Osorio+2017,Sanna+2019}, as well as in H{\sc ii} regions \citep{Meng+2019,Dewangan2020a,Dewangan2020b}. These observations can be quantitatively explained using the models described in \citet{Padovani+2015,Padovani+2016} for cosmic-ray acceleration in protostellar shocks and in  \citet{Padovani+2019} for expanding H{\sc ii} regions.

The idea we propose in this article, with the aim to discriminate between the presence of shock-generated cosmic rays or UV photons,  is to focus on sources where both synchrotron emission and high ionisation rates have been observed in the same region. In this way, the possible effect of UV photons can be ruled out incontrovertibly.

To the best of our knowledge, the only source for which 
both synchrotron 
detection and ionisation rate estimates have been carried out 
so far is the OMC-2 FIR 3/FIR 4 system. Located in the Orion Molecular Cloud 2 (OMC-2), at a distance of $\sim 388$~pc \citep{Kounkel+2017}, the far-infrared sources FIR 3
and FIR 4 
are connected each other
through a filamentary structure \citep[e.g.][]{Hacar+2018}. FIR 3 
harbours four compact continuum sources
\citep{Tobin+2019}, including the 
intermediate-mass protostar HOPS 370, which has a bolometric luminosity of $\sim 360$ $L_{\odot}$ \citep{Furlan+2016}.
Six protostellar objects have been identified in the FIR 4 protocluster
\citep[e.g.][]{Lopez-Sepulcre+2013,Tobin+2019}, 
the most luminous being 
HOPS 108 ($\sim37L_\odot$; \citealt{Furlan+2014}).
The peculiarity of the FIR 4 source is due to a molecular composition that is consistent with a chemistry regulated by a high ionisation rate of up to $\sim 10^{-14}$ s$^{-1}$, which, as stated above, is likely due to a dose of energetic particles very similar to that experienced by the protosolar nebula \citep[see e.g.][and references therein]{Ceccarelli+2014,Fontani+2017,Favre+2018}. 

In addition to the abovementioned ionisation rate estimates,
\citet{Osorio+2017} detected non-thermal emission in three knots along the jet generated by HOPS 370. 
\citet{Shimajiri+2008} suggested an interaction between the HOPS 370 jet and the FIR 4 region that triggers the fragmentation of the latter into dusty cores as well as the formation of the cluster. 
This scenario was supported by \citet{Osorio+2017}, who concluded that FIR 4 could fall in the path of these non-thermal knots, which are moving away from 
HOPS 370.
While the impact of the HOPS 370 jet
on HOPS 108 is still a matter of debate, 
\citet{Tobin+2019} found evidence for an 
interaction with the surrounding FIR 4 region.
If so, by superimposing the map of the emission at centimetre wavelengths (see Fig.~1 in \citealt{Osorio+2017}) on that of the molecular emission (see Fig.~1 in \citealt{Fontani+2017}), we can see that the three non-thermal knots fall within the region east of FIR 4, where the ionisation rate has been estimated to be $4\times10^{-14}$~s$^{-1}$ (red contours in Fig.~1 of \citealt{Fontani+2017}). Our intent is to show how the same locally accelerated cosmic-ray flux is able to explain both the enhanced ionisation and the non-thermal emission.

This paper is organised as follows: In Sect.~\ref{sec:model}
we describe the particle acceleration model, 
compute the expected flux of locally accelerated cosmic
rays, and review the basic equations used in the computation of the
cosmic-ray ionisation rate and synchrotron emission;
in Sect.~\ref{sec:fit} we describe the fitting procedure
to compare the model predictions to the 
observations, and in Sect.~\ref{sec:comparison} we show the
best fit; in Sect.~\ref{sec:conclusions} we discuss the implications of our findings and summarise the main conclusions.

\section{Model}\label{sec:model}
In this section we summarise the main equations for the first-order Fermi acceleration mechanism that is invoked by \citet{Padovani+2015,Padovani+2016} to explain the generation of a locally accelerated cosmic-ray flux in protostellar environments. According to the Fermi mechanism, the presence of magnetic fluctuations around a shock ensures a large pitch-angle scattering of the thermal particles of the local medium. As a consequence, these particles cross the shock back and forth with an average energy gain of $\Delta E/E\propto v/U$ in each cycle (e.g. up-down-up stream), where $v$ is the particle speed and $U$ is the jet velocity in the shock reference frame. The higher the turbulent component of the magnetic field, $\delta B$, the more efficient the acceleration process is. In the case of a parallel shock, one can show \citep{Pelletier+2006,Schalchi2009} that
\begin{eqnarray}\label{BdB}
k_{\rm u}\equiv \left(\frac{B}{\delta B}\right)^{2}&=&%
8\left(\frac{\tP}{10^{-4}}\right)^{-1}\left(\frac{U}{500~{\rm km~s^{-1}}}\right)^{-1}\times\\\nonumber
&&\left(\frac{n}{10^{6}~{\rm cm}^{-3}}\right)^{-0.5}\left(\frac{B}{100~\mu{\rm G}}\right)\,.
\end{eqnarray}
Here, $\widetilde P$ is the fraction of ram pressure transferred to thermal particles, $n$ is the volume density, and $k_{\rm u}$ is the upstream diffusion coefficient normalised to the Bohm coefficient, which
is defined as
\be
\kappa_{\rm B}=\frac{\gamma\beta^{2}m_{p}c^{3}}{3eB}\,,
\ee
where $e$ is the elementary charge, $m_{p}$ is the proton mass, $c $ is the light speed, $\gamma$ is the Lorentz factor, and $\beta=\gamma^{-1}\sqrt{\gamma^{2}-1}$. The limit $k_{\rm u}=1$, reached as $\delta B$ approaches $B$, is called the Bohm limit, and the first-order Fermi acceleration becomes effective at its maximum degree. 

\subsection{Timescales}\label{timescales}
\begin{figure}[!b]
\begin{center}
\resizebox{.9\hsize}{!}{\includegraphics{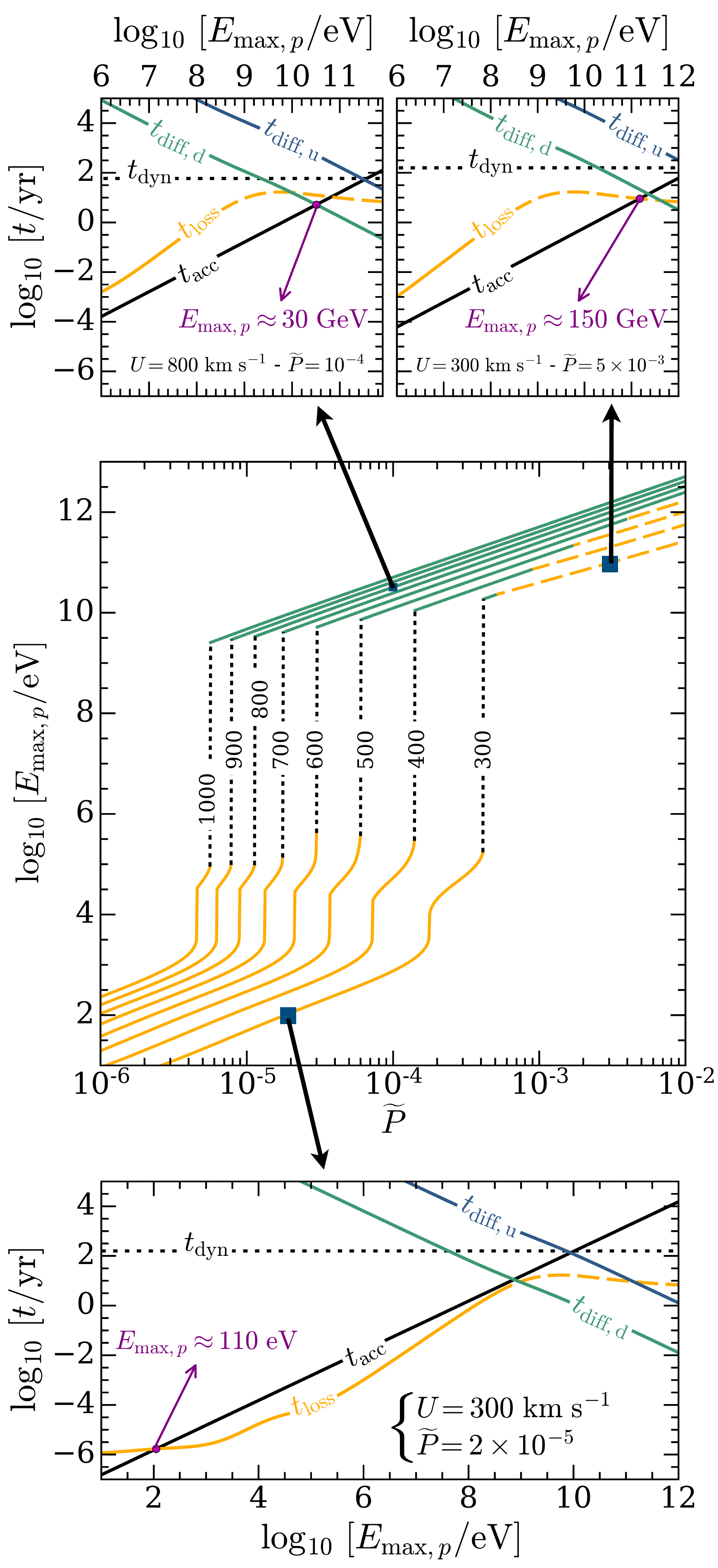}}
\caption{
Comparison of timescales and determination of the maximum energy of protons, $E_{{\rm max},p}$.
Central panel: 
$E_{{\rm max},p}$  versus the shock-accelerated particle pressure, $\widetilde{P}$, normalised to the ram pressure for different jet velocities, $U$, in units of km~s$^{-1}$ (black labels). Solid and dashed orange lines show values of $E_{{\rm max},p}$ constrained by the Coulomb and pion losses, respectively. Solid green lines show $E_{{\rm max},p}$ constrained by the downstream diffusion timescale. The upper and lower panels show the timescales involved in the calculation of $E_{{\rm max},p}$ (Eqs.~(\ref{tacc})-(\ref{tdyn})) for three sets of $U$ and $\widetilde{P}$; $B=100~\mu$G, $n=10^{6}$~cm$^{-3}$, $\ellperp=400$~AU, and $R=12612$~AU.}
\label{sketch-coulomb}
\end{center}
\end{figure}

In the following we summarise the main equations needed to compute the timescales involved in the acceleration process, the maximum energies reached by non-thermal particles, and the corresponding fluxes \citep[for full details and the derivation of the equations, see][]{Drury+1996,Padovani+2015,Padovani+2016,Padovani+2019}.
Protons have to be accelerated before $(i)$ they start to lose energy because of collisions with neutrals, $(ii)$ they escape from the acceleration process, diffusing towards the protostar (upstream) or in the direction perpendicular to the jet (up- and downstream), and $(iii)$ the shock disappears. The associated timescales characterising acceleration, energy losses, up- and downstream escape losses, and the dynamical evolution of the source, are, in unit of years, 
\begin{align}
t_{\rm acc} &= 1.2\times 10^{-2} k_{\rm u}(\gamma-1)\frac{\varrho(\varrho+1)}{\varrho-1}%
        \left(\frac{U}{500~{\rm km~s^{-1}}}\right)^{-2}\left(\frac{B}{100~\mu{\rm G}}\right)^{-1}\,,\label{tacc}\\
t_{\rm loss} &= 10\frac{\gamma-1}{\beta}\left(\frac{n}{10^{6}~{\rm cm^{-3}}}\right)^{-1}%
        \left(\frac{L}{10^{-16}~{\rm eV~cm^{2}}}\right)\,,\label{tloss}\\
t_{\rm diff,u} &= \frac{2.3\times10^{4}}{k_{\rm u}\gamma\beta^{2}}%
        \left(\frac{B}{100~\mu{\rm G}}\right)
        \left(\frac{R}{10^3~{\rm AU}}\right)
        \left[\frac{\min(\ell_\perp,\epsilon R)}{10^3~\rm AU}\right]\,,\label{tdiffu}\\
t_{\rm diff,d} &= \frac{5.7\times10^3}{k_{\rm u}\gamma\beta^{2}}%
        \left(\frac{B}{100~\mu{\rm G}}\right)\left(\frac{\ellperp}{10^{3}~{\rm AU}}\right)^{2}\,,\label{tdiffd}\\
t_{\rm dyn} &= 9.4\left(\frac{R}{10^{3}~{\rm AU}}\right)\left(\frac{U}{500~{\rm km~s^{-1}}}\right)^{-1}\label{tdyn}\,,
\end{align}
where $L$ is the proton energy
loss function \citep[see Sect.~3 and Fig.~1 in][]{Padovani+2018a} and $\varrho$
is the compression ratio defined by
\be\label{compressionratio}
\varrho=\frac{\left(\gamma_{\rm ad}+1\right)M_{s}^{2}}{\left(\gamma_{\rm ad}-1\right)M_{s}^{2}+2}\,,
\ee
where
\be\label{Mson}
M_{s}=\frac{U}{c_{s}}
\ee
is the sonic Mach number,
\be\label{cs} 
c_{\rm s}=9.1\sqrt{\gamma_{\rm ad}(1+x)\left(\frac{T}{10^{4}~{\rm K}}\right)}~\mathrm{km~s^{-1}}
\ee
is the sound speed, $x=n_i/(n_n+n_i)$ is the ionisation fraction ($n_i$ and $n_n$ are the ion and neutral volume densities, respectively), and $T$ is the upstream temperature.
In the following, the adiabatic index is set to $\gamma_{\rm ad}=5/3$. In Eqs.~(\ref{tdiffu}) and (\ref{tdiffd}), $R$ and $\ellperp$ 
represent the distance from the protostar (also known as the shock radius) and the transverse jet width, respectively.
The upstream diffusion timescale (Eq.~(\ref{tdiffu})) is obtained by assuming that the diffusion length in the upstream medium is 
equal to the minimum between $\ell_\perp$ and 
a fraction of the shock radius,
$\epsilon R$ (typically $\epsilon=0.1$), 
that is, $k_{\rm u}\kappa_{\rm B}/U=\min(\ell_\perp,\epsilon R)$.
Since the up- and downstream loss timescales (Eq.~(\ref{tloss})) differ in terms of density ($n$ is a factor $r$ higher downstream) and the Coulomb component of the energy loss function, we evaluated the mean loss timescale
by averaging it over the particle’s up- and downstream residence times
\citep{Parizot+2006}:
\be
\langle t_{\rm loss}\rangle=\left(\frac{t_{\rm loss,u}^{-1}+rt_{\rm loss,d}^{-1}}{1+r}\right)^{-1}\,.
\ee
With the exception of 
$t_{\rm dyn}$, each timescale is a function of the particle energy through
the Lorentz factor and $\beta$.
Thus, by equating the acceleration timescale to the shorter timescales of
loss, up- and downstream diffusion, and dynamical timescales, that is 
\be
t_{\rm acc}=\min(t_{\rm loss},t_{\rm diff,u},t_{\rm diff,d},t_{\rm dyn})\,,
\ee
it is possible to calculate the maximum energy reached by a proton,
$E_{{\rm max},p}$. As shown in the upper and lower panels 
of Fig.~\ref{sketch-coulomb}, this maximum energy extends from non-relativistic to relativistic 
values, depending on the assumptions on the different parameters
($U,\tP,B,n,\ellperp$, and $R$).
More specifically, Fig.~\ref{sketch-coulomb} shows 
$E_{{\rm max},p}$ as a function of the jet velocity and the cosmic-ray pressure, assuming $B=100~\mu$G, $n=10^{6}$~cm$^{-3}$,
$\ellperp=400$~AU, and $R=12612$~AU. 
It is clearly a threshold process: The central panel of the figure shows that the higher $U$ is, the smaller the fraction of ram pressure that needs to be transferred to thermal particles to jump to the relativistic branch is. This can be understood by noticing that the loss timescale is independent of both $U$ and $\tP$, while $t_{\rm acc}\propto \tP^{-1}U^{-3}$ (the dynamical and the upstream diffusion timescales are always too long compared to the other timescales and never determine the maximum energy).
If $\tP$ is low (below $4\times10^{-4}$ and $6\times10^{-6}$ for $U=300$ and $1000$~km~s$^{-1}$, respectively), the maximum energy is determined by the intersection of the acceleration and the loss timescales. In this case, Coulomb losses keep $E_{{\rm max},p}$ very low (see the bottom panel) and the acceleration process is inefficient. At larger $\tP$, depending on $U$, thermal particles are boosted up to relativistic energies. In this case, $E_{{\rm max},p}$ can be constrained either by pion losses (e.g. $U=300$~km~s$^{-1}$ and $\tP>5\times10^{-4}$; see the top right panel) or by the downstream diffusion timescale (e.g. $U>700$~km~s$^{-1}$; see the top left panel). In fact, for high $U$, $t_{\rm acc}$ rapidly decreases, intersecting $t_{\rm diff,d}$, and $E_{{\rm max},p}$ increases with increasing $\tP$ and $U$ since $t_{\rm diff,d}\propto\tP U$.

For application to the HOPS 370 jet, the distances, $R$, of the three synchrotron knots, VLA 12N, 12C, and 12S, from HOPS 370 are about 6955~AU, 10572~AU, and 12612~AU, respectively \citep{Osorio+2017}, while the average size of the synchrotron emitting regions is $\ellperp\sim400$~AU \citep{Tobin+2019}. As a consequence, the upstream diffusion timescale never determines the maximum energy since $t_{\rm diff,u}/t_{\rm diff,d}\simeq 4R/\ellperp\gg1$.

\subsection{Emerging cosmic-ray energy distribution}\label{spectra}
The energy distribution per unit density (hereafter distribution) of shock-accelerated protons is given by
\be\label{Nshp}
\mathscr{N}_p(E)=4\pi p^{2}f(p) \frac{\ud p}{\ud E}\,,
\ee
where $f(p)$ is the momentum distribution at the shock surface.
In the test-particle regime, which is valid in this study, this is a power law,
\be\label{fp}
f(p)=f_{0}\left(\frac{p}{p_{\rm inj}}\right)^{-q}\,,
\ee
with $q=3\varrho/(\varrho-1)$ and $p_{\rm inj}<p<p_{\rm max}$, where $p_{\rm inj}$ is the injection momentum, namely the momentum at which the emerging flux intersects the Maxwellian distribution of thermal particles \citep[see][]{Padovani+2019} and $p_{\rm max}$ directly derives from $E_{{\rm max},p}$.
The normalisation constant, $f_0$, is
\be\label{f0}
f_0=\frac{3}{4\pi}\frac{n\widetilde P}{\mathscr{I}}\left(\frac{U}{c}\right)^2(m_pc)^{q-3}p_{\rm inj}^{-q}\,,
\ee
where
\be\label{I2}
{\mathscr I}=\int_{\widetilde{p}_{\rm inj}}^{\widetilde{p}_{\rm max}}\frac{p^{\,4-q}}{\sqrt{p^{\,2}+1}}\ud p
\ee
and 
$\widetilde{p}_{\rm inj}$ and $\widetilde{p}_{\rm max}$ are the injection and maximum
momenta normalised to $m_{p}c$ \citep[see][]{Padovani+2016}.

The electron injection process in shock acceleration is poorly understood.
Following \citet{Padovani+2019}, we adopted the model of \citet{BerezhkoKsenofontov2000} 
to estimate 
the distribution 
of shock-accelerated electrons,
$\mathscr{N}_e$.
Assuming the same energy of the injected 
electrons as that for protons, namely $p_{{\rm inj},e}=\sqrt{m_e/m_p}p_{{\rm inj},p}$
($m_e$ is the electron mass), at relativistic energies it follows that
\be\label{NeNp}
\frac{\mathscr N_e}{\mathscr N_p}=\left(\frac{m_e}{m_p}\right)^{(q-3)/2}\,.
\ee
The electron maximum energy, $E_{{\rm max},e}$, is limited by synchrotron and inverse Compton losses. However, we 
checked that, with a local radiation field characterised by
an energy density $0.1$--$10$ times that of the interstellar radiation field, as implied by the observed molecular abundance ratios \citep[][]{Karska+2018}, inverse Compton losses are at least one order of magnitude smaller than 
synchrotron losses. Therefore,
$E_{{\rm max},e}$
is obtained by equating the acceleration timescale 
(Eq.~(\ref{tacc})) to the synchrotron timescale, $t_{\rm syn}$, given by
\be\label{tsyn}
t_{\rm syn}=2.7\times10^{9}%
\frac{\gamma-1}{\gamma^2}\left(\frac{B}{100~\mu{\rm G}}\right)^{-2}~\mathrm{yr}\,.
\ee
If $t_{\rm syn}>t_{\rm acc}$ at any energy, then we set $E_{{\rm max},e}=E_{{\rm max},p}$. Finally, at energies larger than $E^*$, where the condition $t_{\rm syn}(E^*)<t_{\rm dyn}$ is fulfilled, the slope of the electron 
distribution, 
$s$, is modified from $\mathscr{N}_e(E)\propto E^s$ to $\mathscr{N}_e(E)\propto E^{s-1}$ \citep{BlumenthalGould1970}.

The fluxes for the HOPS 370 jet knots (number of particles per unit energy, time, area, and solid angle) were computed from the corresponding 
distributions 
as
\be\label{jfromN}
j_k=\frac{\beta_k c}{4\pi}\mathscr{N}_k\,,
\ee
where $k=p,e$.

\subsection{Ionisation rate of locally accelerated particles}\label{ionrate}

\citet{Fontani+2017} found an ionisation rate equal to $4\times10^{-14}$~s$^{-1}$ in a region with an average radius $R_{\rm ion}\sim5000$~AU towards FIR 4. 
Since the angular resolution of these observations is $9.5^{\prime\prime}\times6.1^{\prime\prime}$, corresponding to $3800~{\rm AU}\times2440~{\rm AU}$ at a distance of $\sim400$~pc, we assumed that the three radio knots are located at the centre of the region with the enhanced ionisation rate and that each of them contribute one-third of the total ionisation rate. Ionisation by both primary and secondary electrons is negligible, and in the following we only considered the proton component for the calculation of the ionisation rate (see the next section).  
We computed the propagation of the proton flux at the shock, $j_{p}^{\rm sh}$, in each shell of radius $r$ from $\ellperp/2$ to $R_{\rm ion}$, accounting for the attenuation of the flux according to the continuous slowing-down approximation \citep{Padovani+2009}. Then, the flux in each shell is given by
\be\label{shellflux}
j_{p}(E,r,\delta)=j_{p}^{\rm sh}(E_{0})\frac{L(E_{0})}{L(E)}%
        \left(\frac{{\ellperp/2}}{{\ellperp/2}+r}\right)^{-\delta}\,.
\ee
The kinetic energy of a proton decreases from $E_{0}$ to $E$ after passing through a column density 
\be
N_{r}=nr=\mathcal{R}(E_{0})-\mathcal{R}(E)\,,
\ee
where $\mathcal{R}$ is the proton range function defined as
\be
\mathcal{R}(E)=\int_{0}^{E}\frac{\ud E^{\prime}}{L(E^{\prime})}\,.
\ee
The last factor on the right-hand side of Eq.~(\ref{shellflux}) accounts for the two limiting cases of
diffusion \citep[$\delta\,=\,1$;][]{Aharonian2004} or geometric dilution ($\delta\,=\,2$). Finally, we computed the mean proton flux averaging over the volume of the spherical shell,
\be\label{averagejp}
\langle j_{p}(E,\delta)\rangle=\left(\frac{4\pi}{3} R_{\rm ion}^{3}\right)^{-1}%
        \int_{0}^{R_{\rm ion}}4\pi r^{2}j_{p}(E,r,\delta)\ud r
,\ee
and the corresponding ionisation rate is
\be\label{zetaion}
\zeta_{\delta}=2\pi\int\langle j_{p}(E,\delta)\rangle\sigma_{\rm ion}(E)\ud E\,,
\ee
where $\sigma_{\rm ion}$ is the ionisation cross-section for protons colliding with
molecular hydrogen \citep[see e.g.][]{Rudd+1992}.

\subsection{Synchrotron emission}
\label{sec:synchrotron}

From the shock-accelerated electron distribution, we computed the expected synchrotron flux density, $S_{\nu}$, at a frequency $\nu$ assuming a Gaussian beam profile and an average size of the synchrotron spots of $\mathcal{L}_{\rm syn}\sim400$~AU \citep{Tobin+2019},
\be\label{Snu}
S_{\nu}=\frac{\pi}{4\ln2}\epsilon_{\nu}\theta_{b}^{2}\mathcal{L}_{\rm syn}\,,
\ee
where $\theta_{b}$ is the beam full size at half maximum and $\epsilon_{\nu}$ is the synchrotron specific emissivity,
\be\label{epsnu}
\epsilon_{\nu} = \frac{1}{4\pi}\int_{m_{e}c^{2}}^{\infty}\mathscr{N}_{e}(E)P_{\nu}^{\rm em}(E)\,\ud E\,.
\ee
Here, $P_{\nu}^{\rm em}(E)$ is the total power per unit frequency emitted at frequency $\nu$ by an electron of energy $E$ \citep[see e.g.][]{Longair2011}. We note 
that we did not account for the attenuation of the electron distribution. In fact, assuming that electrons are created at the centre of a non-thermal knot, they traverse a column density $N_{\rm syn}=n\mathcal{L}_{\rm syn}/2$ to reach the outer boundary of a knot. Assuming an average volume density of $n=10^{6}$~cm$^{-3}$, then $N_{\rm syn}=3\times10^{21}$~cm$^{-2}$. Therefore, only electrons with energy lower than $\sim100$~keV are thermalised \citep[see Fig. 2 in][]{Padovani+2018a}. 
As a consequence, relativistic electrons responsible for synchrotron emission are not attenuated. 
 
In principle, synchrotron self-absorption could occur at low frequencies when the emitting electrons absorb synchrotron photons \cite[see e.g.][]{RybickiLightman86}. We computed the optical depth, $\tau_{\nu}=\kappa_{\nu}\mathcal{L}_{\rm syn}$, where $\kappa_{\nu}$ is the absorption coefficient per unit length at frequency $\nu$ defined by
\be\label{knu}
\kappa_{\nu}=-\frac{c^{2}}{8\pi\nu^{2}}\int_{0}^{\infty}E^{2}\frac{\partial}{\partial E}%
\left[\frac{\mathscr{N}_{e}(E)}{E^{2}}\right]P_{\nu}^{\rm em}(E)\,\ud E\,,
\ee
and we found that even at the lowest observing frequency (6~GHz), $\tau_{\nu}\ll1$; as such, the flux density is straightforwardly determined by Eq.~(\ref{Snu}).

\section{Fitting the model to the synchrotron spectrum}\label{sec:fit}

In this section we compare the predictions of our model with the VLA observations of \citet{Osorio+2017} for the three non-thermal knots inside the jet generated by HOPS 370, namely VLA 12N, 12C, and 12S. 
The observations include the C, X, and K 
bands (at 5\,cm, 3\,cm, and 1.3\,cm, 
respectively); also included in the fit are the observations in the Ka band (at 9\,mm) by \citet{Tobin+2019}, as anticipated by \citet{Osorio+2017}.
There are five free parameters in our model for synchrotron emission in Eq. (\ref{Snu}): the jet velocity in the shock reference frame, $U$; the fraction of ram pressure transferred to thermal particles, $\tP$; the magnetic field strength, $B$; the volume density, \nh; and the transverse jet width, $\ellperp$.

Some caveats should be considered before our model is used to interpret the observations. First, due to the lack of information on magnetic field strength and morphology around the shock, we assumed free-streaming 
propagation to compute the attenuation of the proton flux in each shell (see Eq.~(\ref{shellflux})). However, since charged particles propagate following a spiral path around magnetic field lines, the traversed effective column density can be much larger, especially if field lines are tangled \citep{Padovani+2013}. As a result, the ionisation rate computed with Eq.~(\ref{zetaion}) should be considered as an upper limit.
Second, since the energy of the electrons responsible for synchrotron emission is proportional to $(\nu/B_\perp)^{1/2}$,
where $B_\perp$ is the component of the magnetic field vector
projected on the plane of the sky
\cite[e.g.][]{Longair2011}, different parts of the electron distribution are mapped at any given frequency of observation, depending on the local value of $B$. Since $P_{\nu}^{\rm em}$ is also a function of $B$, it follows that the specific emissivity (Eq.~(\ref{epsnu})) spatially depends on the magnetic field. In our model we had to
assume 
 a constant value for $\epsilon_{\nu}$ along the line of sight, and the flux density computed with Eq.~(\ref{Snu}) should be regarded as an approximation.

\subsection{Bayesian method}

We adopted a Bayesian method to infer the best-fit model parameters. Such an approach is particularly helpful in our case because the number of free parameters is commensurate with the number of data points\footnote{This also implies that the reduced $\chi^2$ is the same as $\chi^2$.}. 
In particular, we used
\begin{equation}
\label{eq:bayes}
P({\bm\theta}|{\bm D}) \propto P({\bm \theta}) P({\bm D}|{\bm \theta})
\end{equation}
to derive the full posterior probability distribution,
$P({\bm \theta}|{\bm D}),$ of the parameter vector ${\bm \theta}$\,=\,($U$, $\tP$, $B$, \nh, $\ellperp$) given the data vector, ${\bm D}$ (the observed synchrotron spectrum). 
This posterior is proportional to the product of the prior, $P({\bm \theta),}$ on all model parameters (the probability of a given model being obtained without knowledge of the data)
and the likelihood, $P({\bm D}|{\bm \theta}),$ that the data are compatible with a model generated by a particular set of parameters. 
We assumed that the data are characterised by Gaussian uncertainties, so the likelihood of a given model is proportional to $\exp(-\chi^2/2)$, with
\begin{equation}
\chi^{2} = \sum_{i} \left(\frac{S_{\nu_i}^{\rm obs}-S_{\nu_i}}{\sigma_{\nu_i}^{\rm obs}} \right)^{2}\,,
\label{eq:chi2}
\end{equation}
where $S_{\nu_i}^{\rm obs}$ is the observed flux density at the $i^{\rm th}$ frequency, $\nu_i$; $S_{\nu_i}$ is the corresponding model flux predicted by Eq. (\ref{Snu}); and $\sigma_{\nu_i}^{\rm obs}$ is the standard deviation.
The best-fit parameter vector, $\bm \theta$, is evaluated by constructing the probability density function (PDF) of a given parameter, weighting each model with the likelihood, and normalising to ensure a total probability of unity. Specifically, this is achieved by marginalisation over other parameters:
\begin{equation}
\label{eq:marginalize}
P({\bm\theta}|{\bm D})\,= \int \,P({\bm \theta},Y|{\bm D})\,dY\,,
\end{equation}
where $Y$ is the list of parameters, excluding the parameter of interest.

In our model, the upstream temperature is fixed at the typical value for protostellar shocks of $10^{4}$~K \citep{Frank+2014}, and we assumed a completely ionised medium\footnote{The case of an incomplete ionised medium is discussed in 
\citet{Padovani+2015,Padovani+2016} and in the appendix of \citet{Padovani+2019}.}, $x=1$ \citep{Araudo+2007}.
The chemical model described in \citet{Ceccarelli+2014} constrains the density in the range $8\times10^{5}\le n/{\rm cm^{-3}} \le 2\times10^{6}$ from the observed abundances of N$_{2}$H$^{+}$ and HCO$^{+}$ towards FIR 4. The range of velocity in high-mass protostellar jets is expected to be between 300 and 1000~km~s$^{-1}$ \citep{Marti+1993,Masque+2012}. 
Since jet velocities are one order of magnitude larger than shock velocities, in the following we assumed that $U$ is equal to the jet velocity. In supernova remnants, the fraction of ram pressure 
transferred to thermal particles, $\tP$, 
is of the order of 10\% \citep{BerezhkoEllison1999}.
Since we expect shocks in star-forming regions to be much less energetic, we let $\widetilde P$ vary between $10^{-6}$ and $10^{-2}$. 
Finally, for the magnetic field strength, $B$, we considered a range between 100~$\mu$G and the upper limit of $\sim$10~mG predicted by magnetohydrodynamic jet simulations \citep{Hartigan+2007}.   
The relevant parameter ranges are uniformly sampled, and they are summarised in Table~\ref{tab:params}. For $\tP$, $B$, and \nh, the parameter space is sampled logarithmically (in 77, 40, and nine intervals, respectively), while for $U$, the sampling is linear (29 intervals). Although the average size of a knot is $\mathcal{L}_{\rm syn}\sim400$~AU \citep{Tobin+2019}, 
the shock region from which synchrotron emission arises may be smaller.
Therefore, we considered four values of $\ellperp$, namely 100\,AU, 200\,AU, 300\,AU, and 400\,AU.

\begin{table}[!h]
\caption{Parameters from model fitting$^a$.}
\resizebox{\linewidth}{!}{
\begin{tabular}{lcccccc}
\toprule\toprule
\multicolumn{1}{c}{Source} &
\multicolumn{1}{c}{$\chi^2$} &
\multicolumn{1}{c}{$U$} &
\multicolumn{1}{c}{$\log_{10}\tP$} &
\multicolumn{1}{c}{$B$} &
\multicolumn{1}{c}{$n$} &
\multicolumn{1}{c}{$\ellperp$} \\
&& 
\multicolumn{1}{c}{[${\rm km~s^{-1}}$]} & 
\multicolumn{1}{c}{} & 
\multicolumn{1}{c}{[$\mu{\rm G}$]} &
\multicolumn{1}{c}{[$10^{6}~{\rm cm^{-3}}$]} &  
\multicolumn{1}{c}{[AU]}\\
\midrule
&& \multicolumn{5}{c}{Ranges of uniform priors} \\
\cmidrule{3-7}
&& [$300,1000$] & [$-6,-2$] & [$10^{2},10^{4}$]  & [$0.8,2$] &  [$100,400$]\\
\cmidrule{3-7}
&& \multicolumn{5}{c}{Posteriors$^b$} \\
\cmidrule{3-7}
VLA 12N & 0.09 & $1000_{-350}^{+0}$ & $-4.70_{-0.61}^{+0.33}$ & 
        $242_{-72}^{+103}$ & $0.82_{-0.00}^{+0.41}$ & $100$\\[4px]
VLA 12C & 1.36 & $1000_{-250}^{+0}$ & $-4.85_{-0.45}^{+0.27}$ & 
        $438_{-92}^{+150}$ & $1.23_{-0.31}^{+0.43}$ & $100$\\[4px]
VLA 12S & 3.03 & $350_{-50}^{+350}$ & $-4.76_{-0.54}^{+0.82}$ & 
        $367_{-124}^{+221}$ & $0.91_{-0.09}^{+0.45}$ & $400$\\
\bottomrule
\end{tabular}
}
{\small
\begin{flushleft}
$^a$~~The temperature is fixed to $10^4$\,K and the ionisation fraction to 1. \\
$^b$~~These values correspond to the mode of the PDF and a percentile range between 16\% and 84\% (i.e. a $\pm 1\sigma$ spread assuming Gaussian errors).
\end{flushleft}
}
\label{tab:params}
\end{table}

For each of the three VLA knots identified by \citet{Osorio+2017}, the parameter space of $\bm \theta$ was explored through a 
grid method, stepping through the ranges given by Table \ref{tab:params}. However, because of constraints on the ionisation rate driven by accelerated particles, and the requirement of non-zero radio flux densities, the calculation of $\chi^2$ proceeded in several steps. First, we fixed a pair of velocity and pressure values, ($U$, $\tP$), and computed the emerging cosmic-ray flux for every pair of ($B$, \nh). From this flux, we then calculated the ionisation rate using  all the parameters ($U$, $\tP$, $B$, \nh) and considered only the pairs ($B$, \nh) that gave a $\zeta$($U$, $\tP$, $B$, \nh) consistent with the observations \citep[e.g.][]{Fontani+2017}. Finally, we required the predicted emerging synchrotron flux to be non-zero. This was accomplished by again examining all pairs of ($B$, \nh) and retaining only those parameter configurations of ($U$, $\tP$, $B$, \nh) that resulted in non-zero synchrotron emission.
This 
exercise resulted in about $8\times 10^5$ different combinations of parameters. For these combinations, we calculated $\chi^2$ as in Eq. (\ref{eq:chi2}) and repeated the procedure for each of the four values of $\ellperp$.

The size of the emitting region, $\ellperp$, was sampled somewhat crudely because the model is not particularly sensitive to this parameter; thus, we performed our analysis separately for the four values of $\ellperp$.
Consequently, the marginalisation in Eq. (\ref{eq:marginalize}) was performed over four parameters, ${\bm \theta}$\,=\,($U$, $\tP$, $B$, \nh), and the process was repeated for each value of $\ellperp$. For each value of $\ellperp$, we associated 
the lowest $\chi^2$ value (the highest likelihood) with the probable best-fit $\ellperp$. Using only the models with the best-fit $\ellperp$, determined outside the marginalisation procedure, we took as the best-fit parameter the mode of the PDF, the most probable value in the PDF given by Eq. (\ref{eq:marginalize}). The best-fit values for $U$, $\tP$, $B$, \nh, and $\ellperp$ are reported in Table \ref{tab:params}.

\subsection{Fitting results}
\label{sec:results}

\begin{figure}[!b]
\begin{center}
\includegraphics[width=1\linewidth]{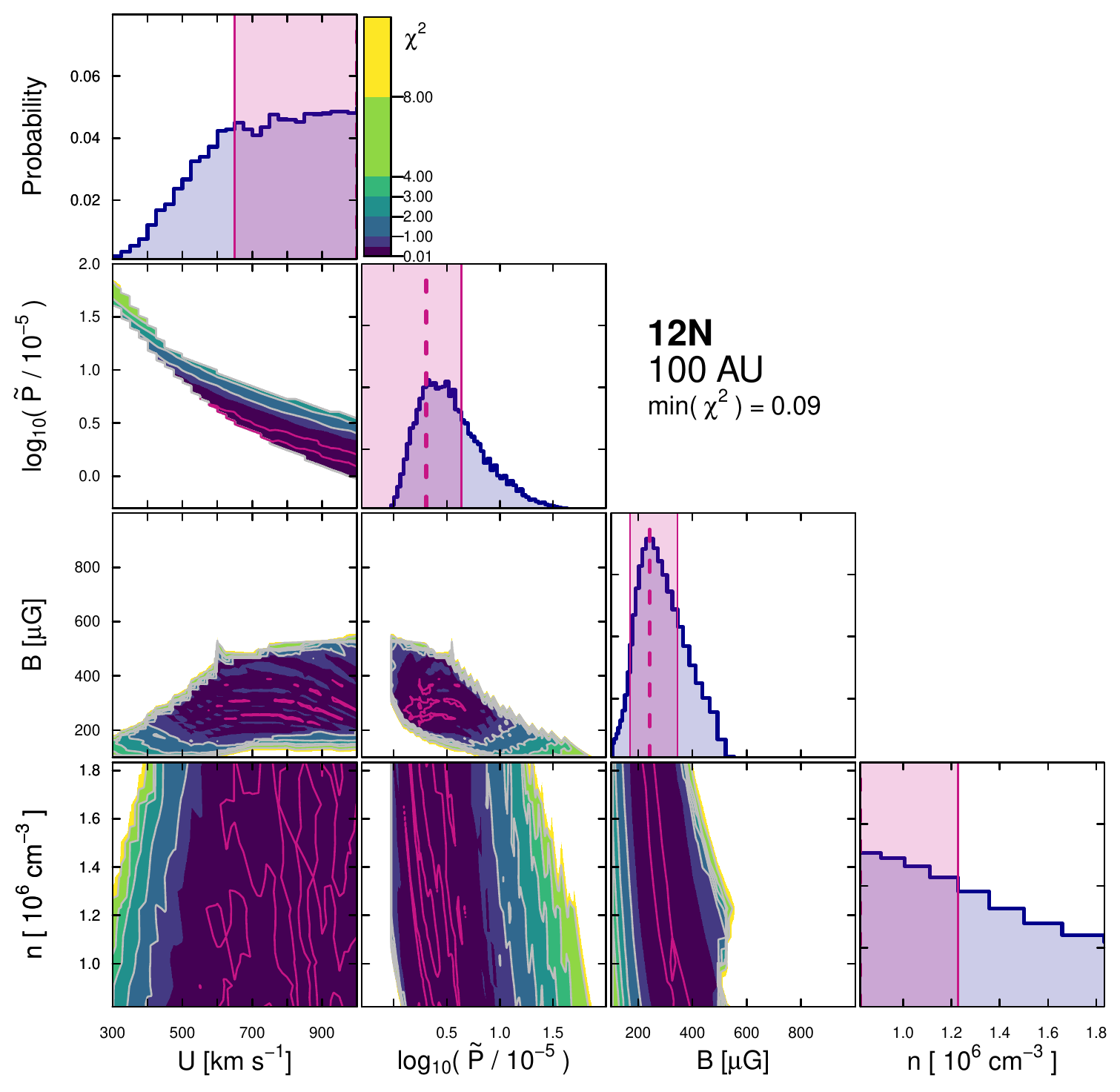}
\caption{Corner plot of the $\chi^2$ surface as a function of the model parameters for VLA 12N.
The region size has been fixed externally for these fits. We show an individual corner plot
for $\ellperp=100$~AU, the best-fit source size. 
The violet contours correspond to the minimum $\chi^2$ value.
The top panels in each column report the probability density distributions for the marginalised parameters;
confidence intervals ($\pm 1\sigma$) are shown as violet shaded rectangular regions, and 
the maximum-likelihood estimate is shown by a vertical dashed line. 
These values together with the upper and lower uncertainties inferred from the confidence intervals
are also reported.
}
\label{fig:vla12n}
\end{center}
\end{figure}

\begin{figure}[]
\begin{center}
\includegraphics[width=1\linewidth]{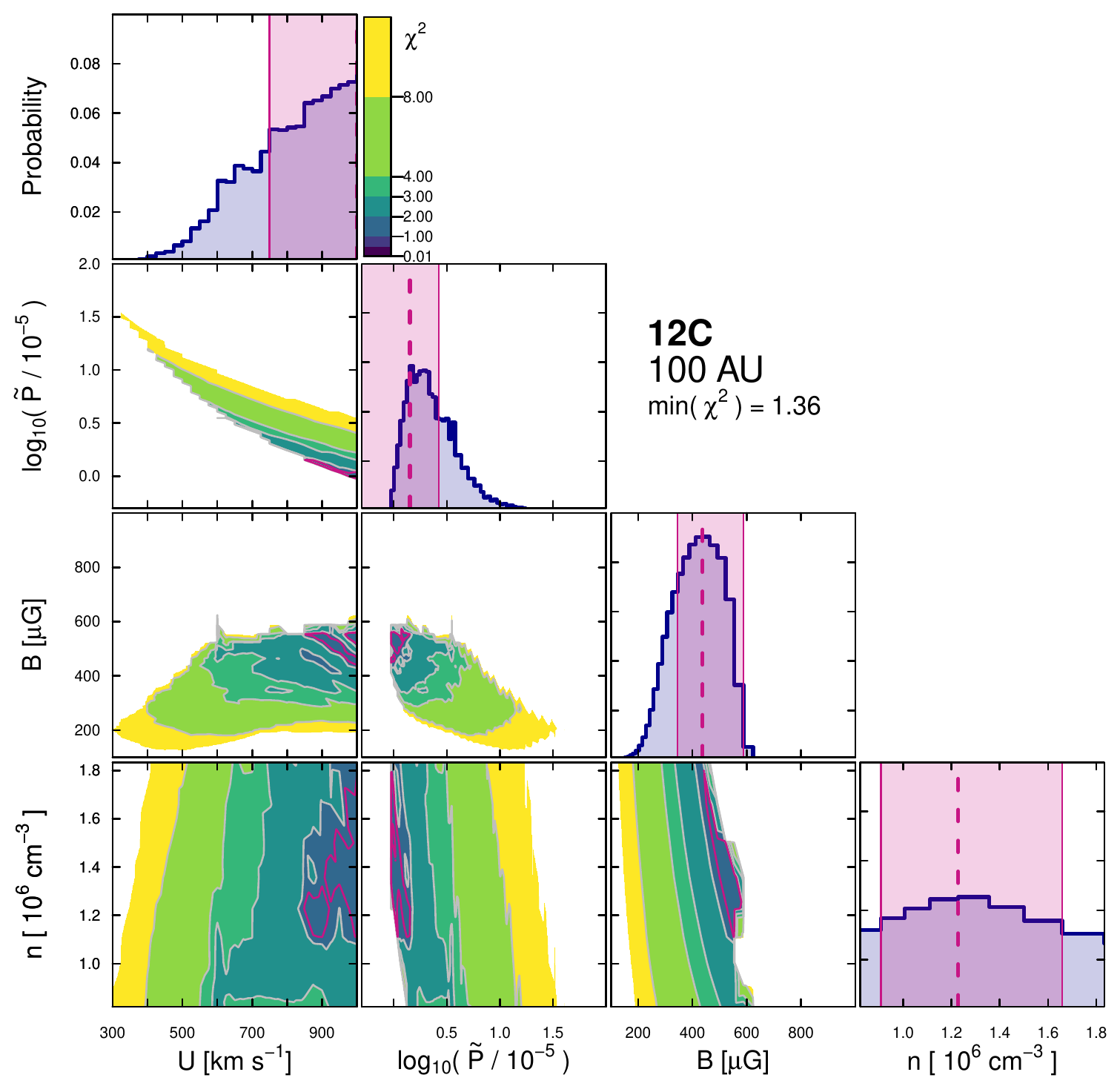}
\caption{Corner plot of the $\chi^2$ surface as a function of the model parameters for the VLA source 12C.
The figure is organised as in Fig. \ref{fig:vla12n}.}
\label{fig:vla12c}
\end{center}
\end{figure}

\begin{figure}[]
\begin{center}
\includegraphics[width=1\linewidth]{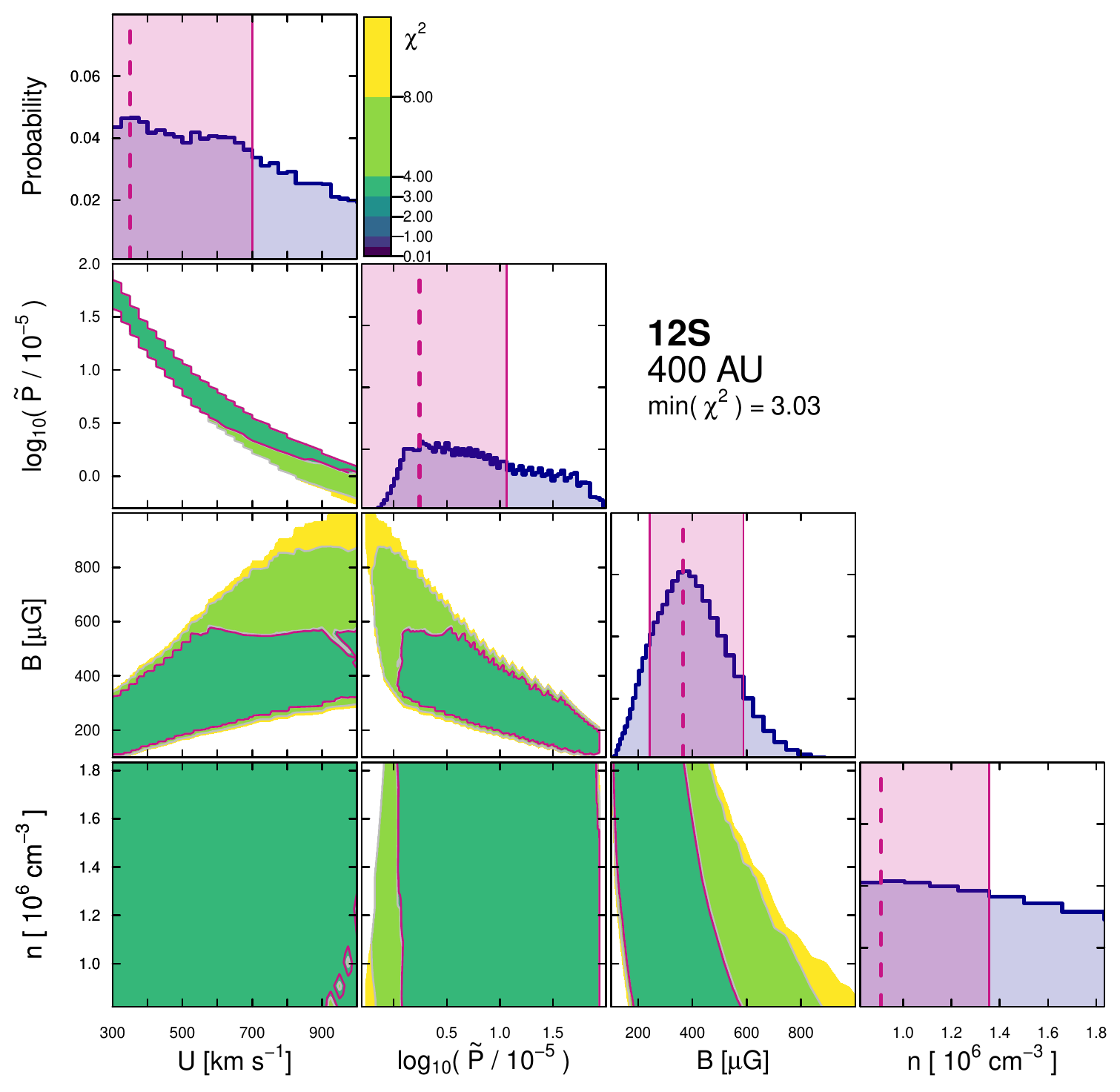}
\caption{Corner plot of the $\chi^2$ surface as a function of the model parameters for the VLA source 12S.
The figure is organised as in Fig. \ref{fig:vla12n}.}
\label{fig:vla12s}
\end{center}
\end{figure}

Figure \ref{fig:vla12n} shows corner plots of the best fit obtained for $\ellperp=100$~AU ($\chi^2=0.09$) for VLA 12N.
We note that the case $\ellperp=200$~AU is virtually indistinguishable in terms of likelihood. 
The remaining transverse jet widths 
for VLA~12N give $\chi^2=1.36$ (300\,AU) and 3.84 (400\,AU).
Figures \ref{fig:vla12c} and \ref{fig:vla12s} show analogous plots for sources VLA~12C and 12S, respectively.
For VLA~12C, the best-fit $\ellperp$ is clearly 100~AU ($\chi^2=1.36$); the other sizes sampled have $\chi^2=3.81$ (200~AU), 8.60 (300~AU), and 13.45 (400~AU).
For VLA~12S, the best fit is $\ellperp=400$~AU  with $\chi^2=3.03$, even if $\ellperp=300$~AU has $\chi^2=3.06$, confirming that the model is not overly sensitive to size in the broad-brush way we have explored it. The other sizes sampled
have $\chi^2=4.39$ (100~AU) and 3.22 (200~AU).
These figures also illustrate that some parameters in our model are mutually correlated.
In particular, the normalised cosmic-ray pressure, $\tP$, and the velocity, $U$, are fairly tightly related, as are $\tP$ and 
the magnetic field strength, $B$.
In contrast, none of the parameters seem to depend strongly on the density, \nh, as shown in the bottom panels of the
figures, although this may be a consequence of the relatively narrow range of densities explored in
our priors because of the observational constraint given by \citet{Ceccarelli+2014}.
Summaries of the knots are given in the following subsections. 

\subsubsection{VLA~12N}
The jet velocity $U$ is large, but it is not well constrained by our choice of parameters
since the best-fit value corresponds to the maximum of the prior, 1000\,\kms.
In contrast, both $\tP$ and $B$ are fairly well determined, both falling towards the
lower end of the range of our expectations in the choice of priors.
Within the tight range of our priors,
the volume density is also somewhat poorly constrained, although
within this range the density was fairly well determined independently by \citet{Ceccarelli+2014}.
The best-fit $\ellperp$ is 100 -- 200\,AU, and it can clearly not be as large as 300\,AU or
400\,AU, at least judging by the significantly larger $\chi^2$ (corresponding to a
likelihood that is four times lower). 
Overall, the fit to the synchrotron spectrum is exceedingly good, with a very low $\chi^2$ value,
possibly indicating that the errors in the radio fluxes are overestimated.

\subsubsection{VLA~12C} The best-fit model parameters are similar to those for source 12N.
Although the best-fit $B$ field is nominally larger, the two values are consistent within the errors.

 \subsubsection{VLA~12S} This source seems to be fundamentally different from the two northernmost ones.
The velocity $U$ is about a factor of three lower than in sources 12N or 12C, as if the jet were decelerating due to an impact
with a denser medium. 
This can be explained by noting that the density of the molecular cloud impacted by the termination shock (in our case, 12S) is 
higher than the density in the jet
\citep[see e.g.][]{Torrelles+1986,Rodriguez-Kamenetzky+2017}. This causes a deceleration of the jet as well as a decrease in the proper motion 
velocity of 12S 
with respect to 12N and 12C as found by \citet{Osorio+2017}.
The 12S proper motion velocity is known, 
namely its bow shock velocity $U_{\rm bs}=37$~km~s$^{-1}$ \citep{Osorio+2017}.
Assuming that the velocity we obtain for 12S 
($U=350$~km~s$^{-1}$) 
represents
the reverse shock velocity, we
can then estimate the jet velocity at the termination shock as 
$\mathbf{U+(\varrho-1)U_{\rm bs}/\varrho}$
\citep[see e.g.][]{Shore2007}.
Since the compression ratio is $\sim4$ (see Eq.~(\ref{compressionratio})), the jet velocity is 378~km~s$^{-1}$.%
\footnote{More specifically, since the shock in 12S turns out to be 
radiative (see Appendix~\ref{adiabaticorradiative}), 
$\varrho$ can be much greater than 1; as such, the jet velocity may approach the upper limit of 387~km~s$^{-1}$.
}
The 12S region itself is also larger, with a size of $\ellperp$ roughly three or four times that of sources 12N and 12C,
revealing a highly collimated jet with an opening angle of $\sim 1.8^{\circ}$,
consistent with the estimate of $2^\circ$ by \citet{Tobin+2019}.

\section{Comparison with observations }\label{sec:comparison}

\begin{figure}[!h]
\begin{center}
\resizebox{1\hsize}{!}{\includegraphics{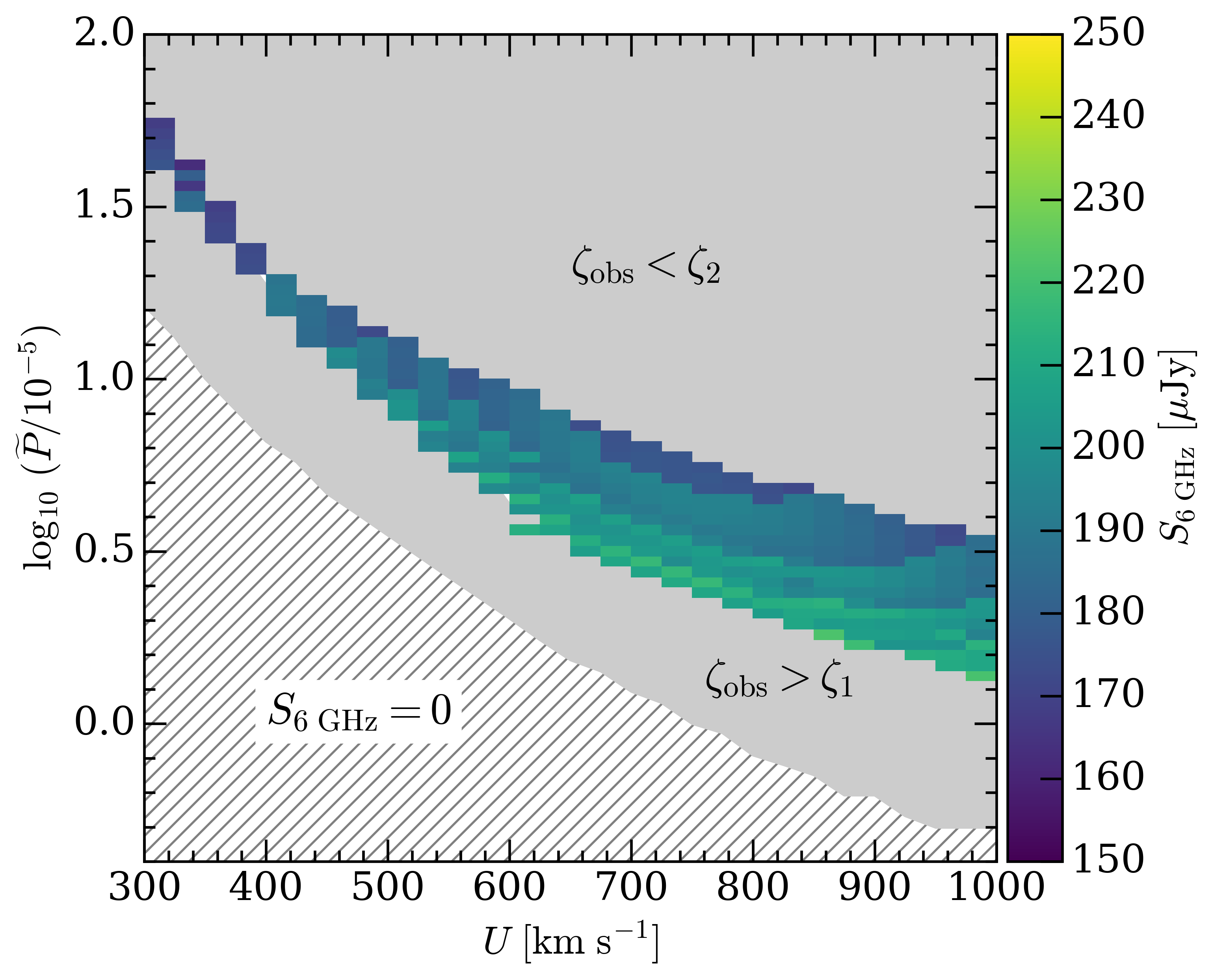}}
\caption{Flux density at 6~GHz obtained with the best-fit procedure for VLA 12C plotted in the space $(\tP,U)$.
The hatched area shows where the expected synchrotron emission is null since the acceleration process is inefficient,
while the two grey areas show the regions where the constraint given by the ionisation rate estimated
by observations, $\zeta_{\rm obs}$, is not satisfied, namely $\zeta_{\rm obs}\notin[\zeta_{2},\zeta_{1}]$.}
\label{example-paramspace}
\end{center}
\end{figure}

Caution should be taken in interpreting VLA observations with our model. The emissivity is a local quantity that depends
on the local electron distribution, $\mathscr N_e$ (Eq.~(\ref{epsnu})). Only with knowing the spatial distribution of both the density and the magnetic
field strength is it possible to accurately compute the attenuation of the electron distribution and to rigorously determine 
the flux density \citep[see e.g.][]{PadovaniGalli2018}. Additionally, the
synchrotron radiation observed is determined by electrons with energy
\be
E_{\rm syn}\simeq1.47\left(\frac{\nu}{\rm GHz}\right)^{1/2}\left(\frac{B_\perp}{100~\mu{\rm G}}\right)^{-1/2}~{\rm GeV}\,
\ee
\citep[see e.g. ][]{Longair2011}.
Since the magnetic field strength is not constant, different energy ranges of the electron distribution are sampled when
integrating along the line of sight. Due to the uncertainty on the spatial distribution of the model parameters
(see Table~\ref{tab:params}), we are forced to assume constant values. However, for this case study the model has a double
constraint, represented by the ionisation rate and the synchrotron emission. This makes the 
parameter space more restricted.
As shown in Fig.~\ref{example-paramspace}, in addition to the region where the acceleration process is not efficient
in extracting charged particles from the thermal pool (such that no synchrotron radiation is emitted), 
two big chunks of the space parameters $(U,\tP)$ are ruled out because
the range of ionisation rates expected by the model does not include the observed
value.\footnote{The ionisation rate depends on $U$ and $\tP$ since
$j_p\propto \tP U^2$; see Eqs.~(\ref{Nshp}), (\ref{f0}), and (\ref{jfromN}).} The extremes of the range are defined by $\zeta_{1}$ and $\zeta_{2}$, the ionisation rates in the case of diffusion and 
geometrical dilution, respectively (see Eq.~(\ref{zetaion})). Thus, a solution is found only if the ionisation rate estimated from 
observations, $\zeta_{\rm obs}$, falls in the interval $[\zeta_{2},\zeta_{1}]$.
Putting together the model estimates for the ionisation rate and the non-thermal emission results in a powerful technique for
isolating the range of validity for the different parameters.

\begin{figure}[!h]
\begin{center}
\resizebox{1\hsize}{!}{\includegraphics{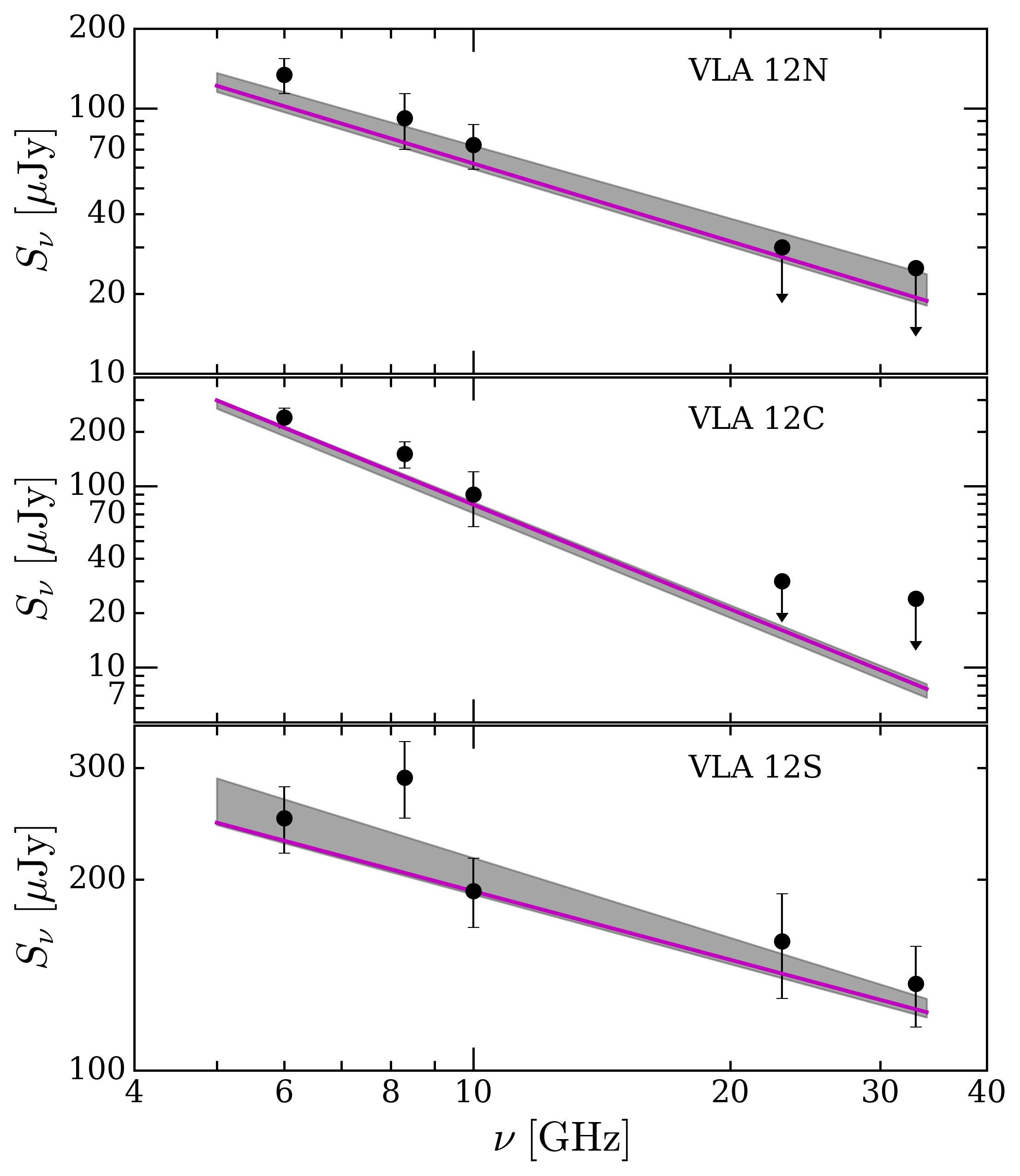}}
\caption{Comparison between the observed flux densities \citep[solid black circles;][]{Osorio+2017,Tobin+2019} and the best-fit models (solid magenta lines)
for the three knots, VLA 12N, 12C, and 12S. The grey shaded areas encompass the models within a confidence interval between
50\% (second quartile) and 96\% (2$\sigma$).
}
\label{vlacomparison}
\end{center}
\end{figure}

With this in mind, in Fig.~\ref{vlacomparison} we compare the flux densities observed with those expected by the model.
As revealed in advance by the low values of $\chi^{2}$, 
the agreement is particularly good for all three knots. The goodness of fit
is a bit poorer for the southern knot since the observation at 8.3~GHz deviates from the clear negative slope traced 
by the detections at the other four frequencies. The spectral indices for VLA 12N, 12C, and 12S are
$\alpha_{\rm 12N}=-0.95^{+0.04}_{-0.02}$,
$\alpha_{\rm 12C}=-1.91^{+0.02}_{-0.01}$,
and $\alpha_{\rm 12S}=-0.36^{+0.01}_{-0.06}$, respectively, in excellent agreement with those computed by \citet{Osorio+2017}.

\section{Discussion and conclusions}\label{sec:conclusions}

In this article we have presented a model showing that cosmic rays accelerated on the surface of a shock, in particular along a protostellar jet, can simultaneously explain the observed molecular abundance (through a high ionisation rate) and the detection of non-thermal emission. Since the latter is generated uniquely by the electronic component of cosmic rays, this allows us to explain the observations without having to rely on shock-generated UV photons. 

We have applied the model to the star-forming region OMC-2 FIR 3/FIR 4, the only one in which a likely co-presence of high ionisation and synchrotron radiation has been identified to date.
In particular, assuming the interaction between 
the jet generated by the protostar HOPS 370 within FIR 3 and the FIR 4 protocluster, we were able to indirectly constrain parameters such as the jet velocity, 
the magnetic field intensity, and the fraction of ram pressure transferred  from the shock to the accelerated particles.

By means of Bayesian methods we found high jet velocities for the northern and central knots 
(650$-$1000~km~s$^{-1}$ and 750$-$1000~km~s$^{-1}$, respectively), while in the southern knot $U$ decreases to 300$-$700~km~s$^{-1}$
as if the jet were hitting a denser medium. The magnetic field strength is well constrained for the three knots between $\sim250~\mu$G
and $\sim450~\mu$G as well as the normalised cosmic-ray pressure between 1.4 and $2\times10^{-5}$ (even if for VLA 12S the error is quite large). We found transverse widths of the jet between 100~AU (VLA 12N and 12C) and 400~AU (VLA 12S), 
revealing a highly
collimated jet with an opening angle of about 1.8$^{\circ}$, consistent with 
previous estimates \citep{Tobin+2019}.

From these quantities, 
it is also possible to evaluate the intensity of the turbulent magnetic field component, $\delta B$, using Eq.~(\ref{BdB}).
For the three knots, VLA 12N, 12C, and 12S, we find $B/\delta B$ equal to
$7^{+24}_{-5}$, $11^{+22}_{-6}$, and $16^{+49}_{-14}$, respectively, 
deviating significantly from the Bohm regime ($\delta B\approx B$).
From this it follows that the magnetic field should not be completely 
randomised, unlike in the Bohm regime. Consequently, polarisation observations could help confirm the non-thermal nature of this emission since synchrotron emission is
linearly polarised.
The fractional polarisation is $\Pi=(3-3\alpha)/(5-3\alpha)$ \citep[e.g.][]{Longair2011}, and, according to our estimates of the 
spectral index, $\Pi$ should be between 67\% and 81\%. 
The values obtained with the above relation must be considered as upper limits since this relation is only valid if the magnetic field direction 
is constant along the line of sight. However, we have shown that the turbulent component of the magnetic field
is not negligible,
in addition to being fundamental for the efficiency of the acceleration process. 
This causes the polarisation fraction to decrease. 
An additional source of depolarisation is represented by the Faraday rotation 
because of the high electron density at the shock position. This effect is even 
stronger at lower frequencies \citep{Lee+2019}.

Despite the uncertainty regarding the spatial distribution of the 
fundamental parameters of the model and its relative simplicity, it is evident that
the combination of measurements of ionisation rates and synchrotron 
emission 
makes it possible to obtain stringent constraints on the physical characteristics of the sources.
Beyond the modelling of this single 
protostellar source, our study aims above all to represent  an input for the observers: Observations of 
co-spatial continuum emission
at centimetre wavelengths and molecular transitions can open a new window on the study of star-forming regions.
A significant advance in the field is expected in the near future, when  a powerful instrument such as 
the Square Kilometre Array (SKA) 
will complement facilities already operating 
at millimetre to metre wavelengths
at millimetre and centimetre wavelengths, such as 
the Atacama Large Millimetre Array (ALMA), 
NOEMA, 
the Very Large Array (VLA), 
the Giant Metrewave Radio Telescope (GMRT), 
the LOw Frequency ARray (LOFAR), and 
the Karoo Array Telescope (MeerKAT).

%%%%%%%%%%%%%%%%%%%%%%%%%%%%
\begin{acknowledgements}
We thank the referee for her/his careful reading 
of the manuscript and insightful comments.
M.P. and L.K.H. acknowledge funding from the INAF PRIN-SKA 2017 program 1.05.01.88.04
and by the Italian Ministero dell'Istruzione, Universit\`a e Ricerca through the grant Progetti Premiali 2012--iALMA (CUP C52I13000140001).
This work has been supported by the project PRIN-INAF-MAIN-STREAM 2017 ``Protoplanetary disks seen through the eyes of new-generation instruments''.
We also are grateful to Robert Nikutta for enlightening discussions about Bayesian inference.
\end{acknowledgements}

\bibliographystyle{aa} % style aa.bst
\bibliography{mybibliography-bibdesk.bib} % your references Yourfile.bib

\appendix

\section{On the adiabatic or radiative nature of the shocks in the HOPS 370 jet}
\label{adiabaticorradiative}

It is usually assumed that the first-order Fermi acceleration mechanism takes place only in the presence of adiabatic shocks.
In order to establish the adiabatic or radiative nature of the shocks in the HOPS 370 jet, we applied the criterion described by \citet{Blondin+1989}. These
authors compared the thermal cooling length, $d_{\rm cool}$, with the 
jet half width at the shock position, $\ellperp/2$. 
If the ratio of these
quantities, $\lambda=d_{\rm cool}/(\ellperp/2)$, is much larger than 1, the
shock is effectively adiabatic as the shock-heated gas does not have time to 
cool before leaving the working surface. If $\lambda\ll1$, the shock is
fully radiative.\footnote{The cooling length is parameterised by 
\citet{Hartigan+1987} and \citet{Heathcote+1998} for two
different ranges of velocities.}
Using the values of the jet velocity in the shock reference
frame, the density, and the jet width listed in 
Table~\ref{tab:params}, we find that the assumption of
adiabaticity is marginally consistent only for the 12N and
12C knots ($\lambda=1.16^{+0.00}_{-1.04}$ and 
$0.77^{+0.26}_{-0.61}$, respectively), while 12S is
clearly a radiative shock 
($\lambda/10^{-3}=2.30^{+55.8}_{-1.55}$).

However, the common assumption that the first-order Fermi
acceleration mechanism cannot occur in radiative collisionless shocks has to be taken with caution for two
reasons: 
$(i)$ radiative shocks can re-accelerate background cosmic rays as well as previously accelerated 
thermal particles, as is likely the case in evolved supernova remnants \citep{Raymond20}; 
$(ii)$ recent detections of gamma-ray emission in classical novae by the Fermi telescope also point 
to the possibility that the first-order Fermi acceleration
mechanism 
takes place
in fast radiative shocks, where the acceleration timescale is shorter than any  
energy-loss timescale, such as that of
Coulomb losses,
that may quench the acceleration process \citep{Metzger15}. 
Therefore, at this stage, it cannot be
definitely ruled out 
that first-order Fermi acceleration also occurs
in radiative shocks.

%\section{}

\end{document}